\documentclass{article}
\usepackage{amsfonts,amsmath}
\mathchardef\mhyphen="2D 
\usepackage[a4paper]{geometry}
\usepackage{mathtools}
\usepackage{caption}
\usepackage{float}
\usepackage{hyperref}
\usepackage[capitalise]{cleveref}
\usepackage{microtype}
\usepackage{nicematrix}

\def\zo/{$0\mkern2mu\mhyphen1$}

\def\nn/{$n \times n$}

\title{ the Aesthetic Asymptotics of the Mayer Series Coefficients for a
Dimer Gas on a Regular Lattice}
\date{\today}
\author{Paul Federbush\\
Department of Mathematics\\
University of Michigan\\
Ann Arbor, MI, 48109-1043}

\newcommand{\abs}[1]{\left\lvert#1\right\rvert}

\begin{document}

\maketitle
\begin{abstract} 
We conjecture that 
for \textbf{all} regular lattices
$b(n)$ is asymptotically of the form in eq.(A1).
 \begin{equation}\label{A1}
\tag{A1}
     (-1)^{n+1} *  b(n)     \sim   \exp{(  k_{-1} n  +  k_{0} ln(n) + \frac{k_{1}}{n} + \frac{k_{2}}{n^{2}}...)}
\end{equation} 
We restrict testing this to lattices for which we know the first 20 Mayer
series coefficients, the $b(n)$. This includes the infinite number of rectangular
lattices, one for each dimension, the tetrahedral lattice ( in this one case
we know only the first 19 coefficients ), and the (bipartite) body centered
cubic lattices, in dimensions 3 through 7. In this paper we will detail results 
for the rectangular lattices in dimensions 2,3,5,11,and 20, for the
tetrahedral lattice, and for the body centered cubic lattices in dimensions
3,4, and 5. These are all bipartite, unfortunately we do not
have an example of a non-bipartite regular lattice for which we know enough
of the $b(n)$ to work with. For the triangular lattice, regular and non-bipartite,
we know the first 14 $b(n)$.
We feel this is not enough terms to make any judgement, hopefully someone
may compute more terms. We work with an 'approximation' that keeps the
first four terms, in $k_{-1}, k_{0}, k_{1}, k_{2}$, in the exponent in eq.(A1).
Agreement will be striking. 

At the end of Part 1 there is a digression on a conjecture in line
with recent applications of the renormalization group to study phase
transitions and the ideas of Cardy,\cite{23-10}. 

In Part 7 there is some study of susceptibility series
for the Ising model on the 2d rectangular lattice, triangular lattice, and honeycomb
lattice; where there is surprising similarity to 
Mayer series on regular graphs, as studied herein.

Also in Part 7 we show, mirabile dictu, that the number of partitions function.
$p(n)$, has the 'magic' property. As does the ith prime function,
actually ithprime(n-1).

\end{abstract}
\Large {PART 1, RESULTS}
\\
\\
In this paper we present an 'approximation' to the right side of eq.(A1) with
$k_{-1}, k_{0}, k_{1}, k_{2}$ present, and chosen in an 'optimal' way for the number of $n$ 
we work with. We put quotes around approximation since we are using a language
only appropriate if the conjecture were true. We begin by presenting the results, for
each lattice we treat, the values of $k_{i}$, $-1 \leq i \leq 2$, and a representative
set of ratios. The rectangular lattices we treat will be designated by their dimension $d$.
the body centered cubic lattices as bcc3, bcc4, and bcc5, and the tetrahedral lattice as th.

We consider one of our regular graphs. Its Mayer series coefficients we
denote by $b(i)$. ( See the discussion in Part 2 to see how these are
all obtained from data in \cite{1-1}. ) We let $B(i)$ be an approximation of the $b(i)$
obtained from eq.(A1) keeping four $k_{j}$, determined below.
We associate 1 to the ratio $b(12)$/$b(8)$, 2 to the ratio $b(16)$/$b(12)$,
3 to the ratio $b(20)$/$b(16)$ ( $b(19)$/$b(15)$ for
th). For $j$ = 1, 2, 3 we let $Q_{j}$ be these ratios. And $q(j)$ the
similar ratios obtained using the $B(i)$,
Notice all of these
values would be the same if the $b(n)$ grew purely exponentially.
In addition to comparing these three ratios to see how well the
$B(i)$ approximate the $b(i)$, we look at what should serve as a more traditional
measure of error to a mathematician, which we call the $\ell_\infty$ error.

\begin{equation*}
\ell_\infty err = \max_{ 8 < i \leq M} {\frac {\abs{(b(i)/b(i-1)-B(i)/B(i-1)}}{\abs{b(i)/b(i-1)}}}
\end{equation*}

We here emphasize that for each lattice the 'optimal' choice of the $k_{i}$
we make DEPENDS ONLY ON the values of the $b(i)$ for that lattice
with $13 \leq i \leq 20$ ( $12 \leq i \leq 19$ for th ). We have put the word 
optimal in quotes because,
although as we will explain later, we have chosen the $k_{i}$ in a way that
would seem theoretically best possible, we have not proven it so.

We first present our results for the rectangular lattices. Table 1 presents
our selection for the 'optimal' choice of the $k_{i}$. All of our computations
have been performed in Maple to 25 digit accuracy. Our table here has
these results reduced to 5 digit accuracy.

\begin{center}
	\begin{table}[H]
		\caption{} \label{table:1}
		\begin{tabular}{|l|l|l|l|l|l|}
			\hline
			$d \to $ & 2 & 3 & 5 & 11 & 20 \\
			\hline
			\hline
			$k_{-1}$&$2.4195$&$2.956$&$3.5689$&$4.4284$&$5.0492$ \\ \hline
			$k_{0}$&$-1.8347$&$-2.081$&$-2.3622$&$-2.4923$&$-2.4992$ \\ \hline
			$k_{1}$&$-.11190$&$-.3009$&$-1.3247$&$-1.6759$&$-1.3711$ \\\hline
		        $k_{2}$&$-.46586$&$-.3162  $&$2.1250$&$3.2049$&$1.4438$ \\\hline

		\end{tabular}
	\end{table}
\end{center}

The values for the representative set of ratios are given in Table 2, along with the
 $\ell_\infty$ error.
 We have been
struck not only by the accuracy, but by the fact that the asymptotic behavior has been so
accurate for such small values of n ( by n=8 certainly ).

\begin{center}
	\begin{table}[H]
		\caption{} \label{table:2}
		\begin{tabular}{|l|l|l|l|l|l|}
			\hline
			$d \to $ & 2 & 3 & 5 & 11 & 20 \\
			\hline
			\hline
	                 $q_{1}$&$7653$&$ 59620$&$ 630720$&$1.8716*10^{7}$&$ 2.2418*10^{8}$ \\ \hline
		         $Q_{1}$&$ 7639$&$ 59590$&$ 635430$&$ 1.8932*10^{7}$&$ 2.2504*10^{8}$ \\ \hline
		         $q_{2}$&$ 9453$&$ 75560$&$ 820150$&$2.4682*10^{7}$&$ 2.9482*10^{8}$ \\ \hline
		         $Q_{2}$&$ 9448$&$ 75560$&$ 821860$&$2.4762*10^{7}$&$ 2.9513*10^{8}$ \\ \hline
		         $q_{3}$&$ 10623$&$ 86160$&$ 948010$&$ 2.8738*10^{7}$&$3.4330*10^{8}$ \\ \hline
		         $Q_{3}$&$ 10621$&$ 86160$&$ 948770$&$ 2.8773*10^{7}$&$ 3.4344*10^{8}$ \\ \hline
                          $\ell_\infty err$&$8*10^{-4}$&$2*10^{-4}$&$3*10^{-3}$&$5*10^{-3}$&$2*10^{-3}$ \\ \hline

		\end{tabular}
	\end{table}
\end{center}

Table 3 and Table 4 tell the corresponding story for the non-rectangular
lattices we treat.

\begin{center}
	\begin{table}[H]
		\caption{} \label{table:3}
		\begin{tabular}{|l|l|l|l|l|}
			\hline
			$ $ & bcc3 & bcc4 & bcc5& th  \\
			\hline
			\hline
			$k_{-1}$ & $3.2884$&$4.0718$&$4.8107$&$2.4649$ \\ \hline
			$k_{0}$ & $-2.0848$&$-2.2375$&$-2.4014$&$-2.0730$ \\ \hline
			$k_{1}$ & $-.042838$&$1.1842$&$-.73884$&$-.67563$ \\\hline
		        $k_{2}$ & $-1.5952$&$-16.748$&$-.66577 $&$3.0761$ \\\hline

		\end{tabular}
	\end{table}
\end{center}

\begin{center}
	\begin{table}[H]
		\caption{} \label{table:4}
		\begin{tabular}{|l|l|l|l|l|}
			\hline
			$ $ & bcc3 & bcc4 & bcc5 & th  \\
			\hline
			\hline
	                 $q_{1}$& $ 2.2506*10^{5}$&$5.2624*10^{6}$&$8.9138*10^{7} $&$8271.5 $\\ \hline
		         $Q_{1}$& $2.2363*10^{5}$&$4.8088*10^{6}$&$8.8808*10^{7} $&$ 8423.5$\\ \hline
		         $q_{2}$& $2.8487*10^{5}$&$6.3872*10^{6}$&$1.1603*10^{8} $&$ 10594$\\ \hline
		         $Q_{2}$& $2.8436*10^{5}$&$6.2209*10^{6}$&$1.1591*10^{8} $&$ 10649$ \\ \hline
		         $q_{3}$& $3.2493*10^{5}$&$7.2517*10^{6}$&$1.3451*10^{8} $&$ 11778$ \\ \hline
		         $Q_{3}$& $3.2471*10^{5}$&$7.1765*10^{6}$&$1.3445*10^{8} $&$ 11807$ \\ \hline
                          $\ell_\infty err$&$3*10^{-3}$&$4*10^{-2}$&$1*10^{-3}$&$7*10^{-3}$  \\ \hline                                                                                        

		\end{tabular}
	\end{table}
\end{center}
The remainder of the paper develops the theory to yield these results.

DIGRESSION Herein we let $d$ be half the number of edges entering each vertex.
We will be concerned with computations motivated by the Lee-Yang Circle
Theorem, \cite{23-10}. In particular we conjecture that different lattices with the
same $d$ value should have nearly equal values of $k_{-1}$, the difference going
to zero as d increases. We present pairs
of lattices and associated pairs of  
$k_{-1}$, corresponding to $d = 2, =4, =8, =16$. Each lattice pair a rectangular
lattice and a non-rectangular lattice. For $d=2$: rect  2.4195,  th  2.4649.  For $d=4$:
rect  3.3087,  bcc3  3.2884.  For $d=8$: rect  4.0893, bcc4 4.0718. For $d=16$:
rect 4.8192, bcc5 4.8107. Not bad, and of course our computations of $k_{-1}$
have error.
\\
\\
\Large {PART 2, RECTANGULAR LATTICES GENERAL THEORY}
\\
\\
We want to describe the general mathematical structure that enabled one to calculate the first 20 Mayer series coefficients $b_d(n)$ \textbf{in every dimension}.
In each dimension, $d$, in addition to the Mayer series coefficients, $b_d(n)$, there is a related series of coefficients $a_d(n)$. 
For each $n$, $a_d(n)$ is a function of $\{b_d(i) | i \leq n\}$ and each $b_d(n)$ is a function of $\{ a_d(i) | i \leq n\}$. 
This setup is detailed simply in \cite{23-9}. 
The $a_d(n)$ are also simply related to the Virial coefficients, by eq.(12) in \cite{1-7}.

The relation between the $\{b_d(n)\}$ and the $\{a_d(n)\}$ was originally developed by a more complicated formalism, \cite{13-3}, Section 5. 
Both formalisms are presented side by side in convenient form in \cite{20-4}. This reference
is a convenient one to use to convert from the $b(i)$ to the $a(i)$ and visa versa. ( A caution, the
same formulas may be used to make similar conversions for the non-rectangular lattices.
BUT then the variable d in the formulas must be taken not to be the physical dimension,
but half the number of edges entering each lattice, it is a regular lattice. Thus for bcc3, $d = 4$. )
The magic relation that enables one to deal with the infinite number of dimensions is the following
\begin{equation}
	a_d(n) = \sum_{\frac{n}{2} - 1 < i \leq n} \dfrac{\alpha_i(n)}{d^i}.
\end{equation}

That such $\alpha$'s exist is proven in \cite{23-8}. It is the more complicated formalism that one works with to prove this formula. (We caution the reader that it is a tough grind to check the details here.)
This equation enables one to find the $a_d(n)$ for all dimensions, for an $n$ value for which the $\alpha_i$ are known; and thus the $b_d(n)$ for such values of $n$. 

Butera and Pernici carried out the truly Herculean task of computing the $b_d(n)$ for $d \leq 10, n \leq 20$. 
From \cite{20-4} or \cite{23-9} one obtained $a_d(n)$ for $d \leq 10, n \leq 20$, and thus the $\alpha_i(n)$ for $n \leq 20$. These results are listed after eq.(6) in \cite{1-1}. Using eq(1) careful mathematical consideration yields
the $a_d(n)$ for $n\leq 20$, all $d$, and thus the $b_d(n)$ for $n\leq 20$, all $d$.

The dimer entropy density, $\lambda_d(p)$, in the form
\begin{equation}
	\lambda_d(p) = \frac{1}{2} \left( p \ln(2d) - p\ln(p) - 2(1-p)\ln(1-p) - p\right) + \sum_{k=2}^\infty a_d(k)p^k
\end{equation}
with $p$ the dimer density, is studied in \cite{1-1,13-3,20-4}, Section 4 of \cite{pc-5,1-7}, and \cite{23-9}. 
A remarkable fact is that $a_d(k)$ is positive for $k \leq 20$ in all dimensions!
`
One can compute using eq.(12) of \cite{1-7} that the first 20 Virial coefficients are also positive in every dimension! 
These two positivities are not directly related, arising as two separate miracles,
the stuff that dreams...of research...are made of/on.
The numerical study of this paper does not involve $\lambda_d(p)$ or the Virial coefficients. 
\\
\\
\Large {PART 3, THE \textit{IDEAL} PROBLEM}
\\
\\
We consider our specific problem, working with $b(n)$ in the range $n \leq 20$,
trying to find the best choice of the $k_{i}$, $-1 \leq i \leq 2$ for the four term
sum in the exponent keeping just these $k_{i}$, to yield the best asymptotic
expression in the range $n \leq 20$. If Plato were alive today working on this
problem he would tell us to solve the five equations:
\begin{equation}
(-1)^{j+1} * b(j)=c * exp( k_{-1} * j + k_{0} * ln(j) + \frac{k_{1}}{j} +  \frac{k_{2}}{j^{2}})     \quad          16 \leq j \leq 20
\end{equation}
for the five variables $\{ c , k_{-1} , k_{0} , k_{1} , k_{2} \}$. But alas Plato doesn't
tell us how to solve this very non-linear set of coupled equations. We hope someone studies
this problem, but we follow another route that results in solving coupled
sets of linear equations!
 \\
\\
\Large {PART 4, THE LINEARIZED PROBLEM}
\\
\\
For a given $ r \geq 1$ we consider the expression
\begin{equation}
\left(c_0+\frac{c_1}{n} +\ldots+ \frac{c_r}{n^{r}}\right)
\end{equation}
and undertake the problem of picking the best values of the $c_{i}$ so that
\begin{equation}
b(n) \approx (-1) * \left(c_0+\frac{c_1}{n} +\ldots+ \frac{c_r}{n^{r}}\right) * b(n-1)
\end{equation}
the approximation becomes more and more exact as n increases.
Id est the left and right side of the equation are equal within an error we neglect
in our calculation. We continue to neglect these errors (which would be very hard
to estimate rigorously) and we will check at the end how well we've done.
We will see that finding such $c_{i}$ for eq.(4) is a \textit{dual} problem
to finding the best $k_{i}$ for eq.(A1). In particular we will have
formulas to go from the set of $c_{i}$ to the set of  $k_{i}$.

In the Appendix we indicate more completely a reformulation of the
asymptotic behavior of the $b(i)$ not in the exponential form given
in the abstract but in the dual form using eq.(5). There should be
some theorems one can prove showing the equivalence of the
two formulations with some assumed conditions, most particularly 
on the convergence properties of the sum in the exponent in eq.(A1).

We turn to our specific problem of seeking an optimal set of the first
four $k_{i}$. We set $r=6$, and later give some explanation of
this choice, selection rather an art than a science. We want eq.(5) to
hold asymptotically, so since we work with a maximum value of
$n =20$ we impose
\begin{equation}
b(n) = (-1) * \left(c_0+\frac{c_1}{n} +\ldots+ \frac{c_r}{n^{r}}\right) * b(n-1), \quad 14 \leq n \leq 20
\end{equation}
Seven linear equations to solve for the seven values of the $c_{i}$, a piece of cake for the computer.
Notice the $c_{i}$ selected are a function of the values of the $b(i)$ for $13 \leq i \leq 20$.

We now follow the line of reasoning to find the values of the $k_{i}$ from the $c_{i}$.
We start with
 \begin{equation}
     (-1)^{n+1} * b(n)      \approx c *   \exp{(  k_{-1} n  +  k_{0} ln(n) + \frac{k_{1}}{n}   ... )}
\end{equation}
Eq.(6) becomes
\begin{equation}
\begin{split}
      b(n)  \approx & c *     \exp{(  k_{-1} n  +  k_{0} ln(n) + \frac{k_{1}}{n}   ... )}
      \approx   ( c_0+\frac{c_1}{n} \ldots+ \frac{c_r}{n^{r} })*\\
      & c*  \exp{(  k_{-1} (n -1) +  k_{0} ln(n-1) + \frac{k_{1}}{n-1}   ... )}
\end{split}
\end{equation}
From which follows
\begin{equation}
 ( c_0+\frac{c_1}{n} \ldots+ \frac{c_r}{n^{r} }) \approx
     \exp{( k_{-1} + k_{0} ln( \frac{n}{n-1}) + k_{1} (\frac {1}{n}-\frac {1}{n-1}) ...)}
\end{equation}
We now carefully collect powers of $1/n$ from the two sides of this equation
\begin{equation}
c_{0}  \approx  exp{(k_{-1})}
\end{equation}
 \begin{equation}
c_{1} \approx    k_{0 } *  exp{(k_{-1})}
\end{equation}
 \begin{equation}
c_{2 } \approx  ( - k_{1 }+ \frac{k_{0}}{2} + \frac {(k_{0})^{2}}{2} ) *  exp{(k_{-1})} 
\end{equation}
 \begin{equation}
k_{-1} \approx ln( c_{0})
\end{equation}
 \begin{equation}
k_{0} \approx  \frac{c_{1}}{c_{0}}
\end{equation}
\begin{equation}
k_{1} \approx  -\frac{c_{2}}{c_{0}} + \frac{c_{1}}{2 c_{0}} + \frac {1}{2} ( \frac{c_{1}}{c_{0}})^{2}
\end{equation}
With some more work we get
\begin{equation}
k_{2}   \approx    -\frac{1}{2} \frac{c_{3}}{c_{0}}+\frac{1}{12}(k_{0}^{3})
     +\frac{1}{4}(k_{0}^{2})
    +\frac{1}{12}(-6 k_{1}+2) k_{0}-\frac{1}{2} k_{1};
\end{equation}
\begin{center}
Table 5 and Table 6 give the values of the $c(i), i \leq 3$.
	\begin{table}[H]
		\caption{} \label{table:5}
		\begin{tabular}{|l|l|l|l|l|l|}
			\hline
			$d \to $ & 2 & 3 & 5 & 11 & 20 \\
			\hline
			\hline
			$c_{0}$ & $11.241$&$19.221$&$35.478$&$83.793$&$155.89$ \\ \hline
			$c_{1}$ & $-20.623$&$-39.991$&$-83.806$&$-208.84$&$-389.61$ \\ \hline
			$c_{2}$ & $9.8647$&$-27.389$&$104.08$&$296.26$&$505.80$ \\\hline
		        $c_{3}$ & $9.8973$&$-5.3245$&$-221.69$&$-772.23$&$-819.22$ \\\hline

		\end{tabular}
	\end{table}
\end{center}

\begin{center}
	\begin{table}[H]
		\caption{} \label{table:6}
		\begin{tabular}{|l|l|l|l|l|}
			\hline
			$ $ & bcc3 & bcc4 & bcc5 & th  \\
			\hline
			\hline
			$c_{0}$ & $26.801$&$58.664$&$122.82$&$11.763$ \\ \hline
			$c_{1}$ & $-55.875$&$-131.26$&$-294.93$&$-24.384$ \\ \hline
			$c_{2}$ & $31.454$&$11.746$&$297.40$&$21.030$ \\\hline
		        $c_{3}$ & $83.406$&$2044.5$&$8.7242$&$-81.212$ \\\hline

		\end{tabular}
	\end{table}
\end{center}
One may want values of $c(i), i > 3$ eventually to estimate errors.
\\
\\
\Large {PART 5, COMPUTATIONS}
\\
\\
We discuss briefly the computations. They were performed in Maple with
25 integer accuracy. I used my modest desktop Mac computer.
A run of the computer program yielded all
the results for a single lattice, the $k_{i}$, the $c_{i}$, the $Q(i)$, the $q(i)$, 
some others we have not recorded here. It ran for a couple hours, the program
between 200 and 250 lines of Maple.When working with the (infinite number
of) rectangular lattices we have stored the values of the $a_{d}(i)$, given 
in [1], (mentioned above ) in the program. Then with the change of a single
input integer, $d$, the dimension of the lattice, one can run the program
to yield results for that lattice. The infinite number of lattices available to us
in one finite program!

To consider a question raised before, I ran this same program with several values
of $r$:1,2, 3,4,5,6. The values of $c_{0}$ and $c_{1}$ appeared to converge
to their values at $r = 6$. With a little thought one sees that these values must
be unique in the asymptotic limit. If one increased $r$ further, 7,...eventually
results will degenerate because one will be using values of $b(i)$ for $i$
far below 20. There will be an optimal value of $r$,which may depend on
the lattice...but I feel 6 is a good choice...picking it an art rather than a science
as we said. Tables 7 and 8 below give the values of $c(0)$ and $c(1)$ respectively
for the rectangular lattices we have addressed, illustrating some of the comments above.
\begin{center}
	\begin{table}[H]
		\caption{} \label{table:7}
		\begin{NiceTabular}{|l|l|l|l|l|l|}
			\hline
			\diagbox{$r$}{$d$}  &  2 & 3 & 5 & 11 & 20 \\
			\hline
			\hline                                                                                                                                                       
 
1 & $ 11.1000$ &$19.0000$&$ 34.8000$&$ 81.9000$&$ 152.000$ \\ \hline 
2 & $ 11.2000$&$ 19.2000$&$ 35.4000$&$ 83.5000$&$ 155.000$ \\ \hline 
3 & $ 11.2400$&$ 19.2300$&$ 35.4600$&$ 83.7000$&$ 155.800$ \\ \hline 
4 & $ 11.2410$&$ 19.2210$&$ 35.4710$&$ 83.7700$&$ 155.870$ \\ \hline 
5 & $ 11.2411$&$ 19.2199$&$ 35.4740$&$ 83.7770$&$ 155.883$ \\ \hline 
6 & $ 11.2408$&$ 19.2200$&$ 35.4750$&$ 83.7820$&$ 155.897$ \\ \hline

		\end{NiceTabular}
	\end{table}
\end{center}
\begin{center}
	\begin{table}[H]
		\caption{} \label{table:8}
		\begin{NiceTabular}{|l|l|l|l|l|l|}
			\hline
			\diagbox{$r$}{$d$}  & 2 & 3 & 5 & 11 & 20 \\
			\hline
			\hline
		        1& $ -18.60$&$ -34.70$&$ -67.70$&$ -165.0$&$ -309.0 $\\ \hline
			2& $ -20.60$&$ -40.00$&$-81.10$&$ -200.0$&$ -376.0$ \\ \hline
			3& $ -20.66$&$-40.14 $&$-83.04 $&$ -206.0$&$ -386.0$ \\ \hline
			4& $ -20.66$&$ -39.99$&$ -83.43$&$ -207.4$&$ -388.0$ \\ \hline
			5& $ -20.65$&$ -39.93$&$ -83.56$&$ -207.9$&$-389.0$ \\ \hline
			6& $ -20.64$&$ -39.93$&$ -83.62$&$ -208.1$&$ -389.1$ \\ \hline

		\end{NiceTabular}
	\end{table}
\end{center}
This is just the edge of possible such numerical studies.
 \\
\\
\Large {PART 6, GENESIS}
\\
\\
For each manifold the $b(i)$ and the $a(i)$ each determine the other.
Thus the magical positivity properties associated to the $a(i)$ are
encoded in the behavior of the $b(i)$. It was hoping to learn something
about the positivities that I began looking at the asymptotic behavior
of the $b(i)$. I have failed in this venture...to this time.

As to the discovery of the asymptotic exponential behavior presented
in the Abstract of this paper, there were a number of weak or
faulty arguments, some lucky guesses, all of whose  recounting
would be pointless. The one glaring clue that trumpeted the
discovery was its truth for the $d =1$ rectangular manifold!
We present the well known exact expression for this case:
 \begin{equation}\label{eq:17}
	 b(n) = 
	 \begin{cases}
		 ~ & \\
		 \hfil 1 &\quad n = 1 \\
		 ~&~ \\ 
		 (-1)^{n+1} \left(\dfrac{1}{n}\right) \left(\dfrac{(2n-1)!}{(n-1)!n!}\right) &\quad n>1. \\
		 ~&~ 
	 \end{cases}
 \end{equation}
This with employment of the Stirling's Series leads to our desired
asymptotic form in $d = 1$.
\\
\\
\Large{PART 7, ODDS AND ENDS}
\\
\\
We note two discoveries we have made in the course of the current
work. Mysterious and suggesting
tough mathematical physics research. Their exploration  should help in 
finding the \textit{meaning} of the current paper.

1) Each of the regular graphs we have dealt with in this paper has a
dimension $d$, defined so that $ 2d$ is the number of edges entering
each vertex. It is this value of $d$ we have used in computing the $b(i)$
from the given $a(i)$ by formulas in \cite{20-4}. We have tried to go through
this computation using other values of $d$ in a sufficient number of
cases that we now conjecture: Everything we did in this paper works
no matter what $d$ one uses! One gets different $b(i)$ and $B(i)$
but the error between them asymptotically vanishes. Why!

2) We have thrown our machinery at the first 20 terms in the three (incredibly long )
series in \cite{11} for the chi-HIGH expansion of the susceptibility of the
Ising model on the three two dimensional lattices, the 2d rectangular lattice,
the triangular lattice, and the honeycomb lattice. With the $a(i)$ treated 
as the $a(i)$ from \cite{1-1}, and simikarly getting $b(i)$ and $B(i)$. 
BUT these $b(i)$ are all positive!  
The agreement
between $b(i)$ and $B(i)$ is even more striking than in the regular graph
cases of this paper. The following table includes these results as well as the
same computations done for the number of partitions function $p(n)$, and
the number of primes function with a displacement, ithprime(n-1). 

\begin{center}
	\begin{table}[H]
		\caption{} \label{table:9}
		\begin{tabular}{|l|l|l|l|l|l|}
			\hline
			$ $ & rect & tri & honey &  $p(n)$ &$ithprime(n-1)$\\
			\hline
			\hline
	                 $q_{1}$& $5.53317 *10^{8}$&$3.46434*10^{9}$&$1.42012*10^{8} $&$2.40313*10^{7}$&$2.40298*10^{7}$\\ \hline
		         $Q_{1}$& $5.53302*10^{8}$&$3.46424*10^{9}$&$1.42008*10^{8} $&$2.40308*10^{7}$&$2.40292*10^{7}$\\ \hline
		         $q_{2}$& $7.45366*10^{8}$&$4.66790*10^{9}$&$1.91350*10^{8} $&$3.23938*10^{7}$&$3.23911*10^{7}$\\ \hline
		         $Q_{2}$& $7.45361*10^{8}$&$4.66786*10^{9}$&$1.91348*10^{8} $&$3.23936*10^{7}$&$3.23909*10^{7}$\\ \hline
		         $q_{3}$& $8.77067*10^{8}$&$5.49320*10^{9}$&$2.25182*10^{8} $&$3.81275*10^{7}$&$3.81239*10^{7}$\\ \hline
		         $Q_{3}$& $8.77065*10^{8}$&$5.49319*10^{9}$&$2.25181*10^{8} $&$3.81274*10^{7}$&$3.81239*10^{7}$\\ \hline
                          $\ell_\infty err$&$4*10^{-6}$&$1*10^{-5}$&$1*10^{-5}$&
                          $3*10^{-7}$&$3*10^{-5}$ \\ \hline                                                                                        

		\end{tabular}
	\end{table}
\end{center}

\textbf{Acknowledgement} We thank David Brydges for a long discussion
of encouragement and direction. We thank Jacques Perk for encouragement
and much information on point 2) in Part 7.
\section*{APPENDIX }
We let $\mathcal{S}$ be
\begin{equation}
\mathcal{S} = \{ r , k , c_{0} , c_{1}, .... c_{r} \}     \nonumber
\end{equation}
with $r$ an integer $\ge 1$, $k$ an integer $\ge 1$, and $c_{i}$
real numbers. For such a set $\mathcal{S}$ we define an approximation
to $b(n)$, $\hat{b}(n)$ = $\hat{b}(n, \mathcal{S})$ by
\begin{equation}\label{B1}
\tag{B1}
\hat{b}(n) = 
\begin{cases}
	b(n) ,&\quad n < k  \\
	f(n), &\quad n\geq k
\end{cases}
\end{equation}
with
\begin{equation}\label{B2}
\tag{B2}
 f(n)=(-1)^{n-k+1} b(k-1)  \prod\limits_{i=k}^{n}\left(c_0+\frac{c_1}{i} +\ldots+ \frac{c_r}{i^{r}}\right)
\end{equation}
Using this definition we state the type of theorem we would like to have true.
\section*{Desired Result}
 Given $d$, (dimension), $r \ge 1$, (degree of approximation), and $\epsilon$, (desired accuracy), there are $k$, 
$c_0$, $c_1$,....$c_r$, all functions of $d$, $r$, $\epsilon$, such that
with approximation $\hat{b}(i)$ defined by these variables as in the abstract, one has
\begin{equation}
\label{B3}
\tag{B3}
	\frac{\abs{b(i)-\hat{b}(i)}}{\abs{b(i)}} \leq \epsilon, \text{ for all } i.
\end{equation}

As was stated before $c_0$ and  $c_1$  will be uniquely determined (if this all makes sense). Of course this 'desired theorem' if true for $ r = 1$ will be true for all
 $r$. But there are obviously desired refinement conjectures that one hopes
 are true, for which the form of the conjecture is different for different values of $r$.


\begin{thebibliography}{1-1}
\bibitem{1-1}P. Butera, P. Federbush and M. Pernici, ``Higher-order expansions for the entropy of a dimer or a monomer-dimer system on $d$-dimensional lattices'', Phys. Rev. E {\bf 87},062113 (2013)
	
\bibitem{bp3-2} P. Butera and M. Pernici, 
  ``High-temperature expansions of the higher susceptibilities for the
  Ising model in general dimension $d$'',  \emph{Phys. Rev.} E {\bf
    86}, 011139 (2012).
    
 \bibitem{13-3} P. Federbush and S. Friedland, ``An Asymptotic Expansion
 and Recursive Inequalities for the Monomer-Dimer Problem",
 \emph{J. Stat. Phys.} {\bf 143}, 306 (2011).
 
 \bibitem{20-4} P. Federbush, ``For the Monomer-Dimer $\lambda_d(p)$, 
the Master Algebraic Conjecture'', arXiv:1209.0987.

 \bibitem{pc-5} Peter Csikvari, ``Matchings in Vertex-Transitive Bipartite
  Graphs'', \emph{Israel Journal of Mathematics} 215 (2016), 99-134.

\bibitem{pc-6} Barry M. McCoy, "Advanced Statistical Mechanics",
 \emph{Oxford University Press} (2010).

\bibitem{1-7}P. Butera, P. Federbush and M. Pernici, ``Positivity of the
Virial Coefficients in lattice dimer models and upper bounds on the number
of matchings on graphs'', Physica A 437 (2015) 278-294.

     \bibitem{23-8} P. Federbush, ``Dimer $\lambda_d$ Expansion, Dimensional
Dependence of $j_n$ Kernels'', arXiv:0806.1941.

    \bibitem{23-9} P. Federbush, ``the Dimer Gas Mayer Series, the Monomer-
Dimer $\lambda_d(p)$, the Federbush Relation'', arXiv:1207.1252.
 \bibitem{23-10} John Cardy, "the Yang-Lee Edge Singularity and Related
 Problems" arXiv:2305.13288.
 
 \bibitem{11} Y. Chan, A. J. Guttmann, B. G. Nickel, J. H. H. Perk, "the Ising
 Susceptibility Scaling Function",  J. Stat. Phys. 145, 549-590, 2011.

\end{thebibliography}
\end{document}